\documentclass[12pt]{article}
\usepackage{amstex}

\renewcommand{\P}{{\cal P}}
\newcommand{\M}{{\cal M}}

\let\dl=\delta
\let\th=\theta
\let\al=\alpha
\let\bt=\beta
\let\gm=\gamma
\let\lm=\lambda

\let\p=\partial
\let\la=\langle
\let\ra=\rangle

\newcommand{\TT}{\mathbf{T}}

\newcommand{\D}{\mathcal{D}}

\newcommand{\xdot}{{\dot x}}

\newcommand{\ket}[1]{|#1\rangle}
\newcommand{\bra}[1]{\langle#1|}
\newcommand{\eq}[1]{eq.~(\ref{#1})}

\renewcommand{\d}{{\rm d}}
\newcommand{\bydef}{\stackrel{\rm def}{=}}

\title{{}\hfill
    \raisebox{2cm}[0pt]{
      \begin{array}[b]{l}
        \mbox{\small IFP/101/UNC}\\
        \mbox{\small\tt hep-th/9609220}
      \end{array}
      }\\
    De Rham Cohomology of the Supermanifolds\\
    and\\
    Superstring BRST Cohomology}

\author{
  Alexander Belopolsky
  \thanks{\tiny Supported in part by the US Department of
    Energy under Grant DE-FG 05-35ER40219/Task~A.}\\
  \em
    Department of Physics and Astronomy\\
    University of North Carolina\\
    Chapel Hill, NC 27599-3255, USA
  }
\date{September 27, 1996} 

\begin{document}
\maketitle
\begin{abstract}
  We show that the BRST operator of Neveu-Schwarz-Ramond superstring
  is closely related to de Rham differential on the moduli space of
  decorated super-Riemann surfaces \(\P\). We develop formalism where
  superstring amplitudes are computed via integration of some
  differential forms over a section of \(\P\) over the super moduli
  space \(\M\). We show that the result of integration does not depend
  on the choice of section when all the states are BRST physical. Our
  approach is based on the geometrical theory of integration on
  supermanifolds of which we give a short review.
\end{abstract}

\section{Introduction and summary}
In this paper we show that the superstring BRST operator is closely
related to the de Rham differential on the moduli space of decorated
super Riemann surfaces. The analogous result for the bosonic string is
well known \cite{ag88-1,cit-zwiebachlong} and has numerous
applications.  Two important applications deserve to be mentioned. The
first is the method for the calculation of string amplitudes with BRST
physical states that do not have primary representatives e.g., the
ghost dilaton \cite{cit-bergmanzwiebach, bz95}. The second is the
relation between the global symmetry group and BRST cohomology at
ghost number one e.g., ghost number one cohomology of the critical
bosonic string in the flat non-compactified background is isomorphic
to the Lie algebra of Poincar\'e group \cite{cit-astbel}.
Generalization of these applications to the superstring will be
presented elsewhere \cite{bel96}.

The paper consists of two parts. The first is devoted to a review of
the geometric theory of integration on supermanifolds.  This includes
differential and pseudodifferential forms, the relation between the two,
their integrals and the de Rham differential. The second part contains
the main result: an interpretation of the superstring measure as a
differential form and a relation between the BRST operator and the
de Rham differential.

\section{Differential forms on supermanifolds}
\label{sec:dfs}
A common misconception in the theory of supermanifolds is that
differential forms have nothing to do with the theory of integration.
Indeed, a na\"\i{}ve generalization of differential forms to the case
of a supermanifold with even coordinates \(x^a\) and odd coordinates
\(\xi^\al\) leads to functions \(F(x,\xi; \d x,\d\xi)\) that are
homogeneous polynomials in \((\d x,\d\xi)\) (note that \(\d x\) is
Grassmann odd and \(\d\xi\) is Grassmann even) and we will see shortly
that such forms cannot be integrated over supermanifolds.

In the pure even case the degree of the form can only be less or equal
than the dimension of the manifold and the forms of the top degree
transform as measures under smooth, orientation preserving coordinate
transformations. This allows one to integrate the forms of the top
degree over the oriented manifolds and forms of degree \(k\) over the
oriented subspaces. On the other hand, forms on a supermanifold may
have arbitrary large degree due to the presence of commuting
\(\d\xi^\al\) and none of them transforms as a Berezinian measure.

The correct generalization of the differential forms that can be
integrated over supermanifold was first suggested by Bernstein and
Leites \cite{bl77}.  {\em Pseudodifferential forms\/} of Bernstein and
Leites, which generalize the notion of inhomogeneous differential forms
(formal linear combinations of differential forms of different degree)
are defined as {\em arbitrary generalized\/} functions \(F(x,\xi; \d
x,\d\xi)\) or distributions on \(\hat M=\Pi TM\), where \(M\) is the
manifold and \(\Pi\) indicates that the parity was changed to the
opposite in the fibers of the tangent bundle.

We will often combine even and odd coordinates into one symbol \(x\).
Berezin integral of a pseudoform \(\omega(x,\d x)\) over \(\hat M\)
does not depend on the choice of coordinates (two Beresinians, one
from the change of \(x\), the other from the change of \(\d x\)
cancel each other due to the parity change). We define an integral of
a pseudoform \(\omega\) over the manifold \(M\) as follows
\begin{equation}
  \label{intdef}
  \int_M \omega = \int_{\hat M}\D(x)\D(\d x)\,\omega(x,\d x),
\end{equation}
where the integral on the right hand side is an ordinary Berezin
integral. Now we can see that the integrals of pseudoforms that are
polynomial in \(\d x\) diverge unless the odd dimension of the
supermanifold is zero. Pseudoforms can be integrated over arbitrary
submanifolds by reducing them to the submanifold and applying
\eq{intdef} with \(M\) replaced with the submanifold. In the case of a
supermanifold with zero odd degree our integration procedure
reproduces an ordinary integral of an inhomogeneous differential form
which extracts the homogeneous component of the proper degree and
integrates it over the submanifold.

The de Rham differential \(\d\) can be defined on the pseudoforms on
\(M\) simply as
\begin{equation}
  \label{derhamd}
  \d\omega(x,\d x) = \sum_{A=1}^{\dim\,M} \d x^A \,
  {\p\omega\over\p x^A}(x,\d x),
\end{equation}
where \(A\) is a {\em generalized index}. For \(\dim\, M = n|m\),
index \(A\) takes \(n\) even and \(m\) odd values.

Stokes theorem is valid for the pseudoforms, yet some caution is
necessary while evaluating the Berezin integral over a bounded domain
(for details, see ref.~\cite[page 21]{ThVor91}).

The major drawback of the theory of pseudoforms is the lack of
grading, and thus the cohomology. This problem was alleviated in the
geometric integration theory developed by Th.~Voronov and A.V.~Zorich
(see ref.~\cite{ThVor91} and references therein). In this
theory (homogeneous) forms on the supermanifold are presented as
generalized Lagrangians of a certain type.  Forms of any even and odd
degree can be extracted from a pseudoform very much like homogeneous
components can be extracted from inhomogeneous differential forms.
This is achieved with the {\em Baranov-Schwartz transformation\/}
\cite{BSchw84, vz86},
\begin{equation}
  \label{bschw}
  \lambda^{r|s}: \omega\mapsto L^{r|s}_\omega,\quad
  L^{r|s}_\omega(x,\xdot)\bydef
  \int \D(\d t)\,\omega\big(x,\sum_{F=1}^{r|s}\d t^F\xdot_F^A\big).
\end{equation}
Forms of degree \(r|s\) can be integrated over non-singular
subvarieties of dimension \(r|s\) and a de Rham differential is defined
so that it maps \(r|s\)-forms to \(r+1|s\)-forms . Both integral and
differential commute with the Baranov-Schwartz transformation.

\section{Superstring measure as a differential form}
First, let us recall some basic facts about the operator formalism for
the superstring. String states are vectors in the tensor product of a
state space of a superconformal field theory and the Fock space of
superconformal ghosts. The latter is generated by two supeconformal
fields \(B(z,\th)=\bt(z)+\th\,b(z)\) and
\(C(z,\th)=c(z)+\th\,\gm(z)\). 

The operator formalism associates a bra state \(\bra{\Sigma}\) with any
punctured super Riemann surface decorated by a choice of local 
coordinates around each puncture.  This state is to be saturated by
a number of Neveu-Schwarz and Ramond string states equal to the
corresponding number of punctures on \(\Sigma\). 

Given a local superconformal vector field \(V^{(i)}(z,\theta)\) for
each puncture on \(\Sigma\), one can define a Schiffer variation
\(\dl_V\Sigma\). Under a Schiffer variation the bra state
\(\bra{\Sigma}\) transforms as follows
\begin{equation}
  \label{schrif}
  \dl_V\bra\Sigma=\bra\Sigma \la \TT, V \ra,
\end{equation}
where \(\TT(z,\th) = {1\over2} G(z)+\th\, T(z)\) is the total
superconformal energy-momentum tensor and the pairing \(\la \TT, V
\ra\) is given by
\begin{equation}
  \label{pairing}
  \la \TT, V \ra=\sum_{\rm punctures}\oint \d z \d\th\,
  \TT^{(i)}(z,\th) V^{(i)}(z,\th).
\end{equation}
Another important property of \(\bra\Sigma\) is that \(\bra\Sigma Q
=0\) where \(Q=\sum Q^{(i)}\) is the sum of BRST operators, one for
each puncture.

Let \(\ket{\psi_1},\ket{\psi_2}\cdots\ket{\psi_n}\) be string states
that saturate \(\bra{\Sigma}\) and \(\ket\Psi\) be their tensor
product.  The string amplitude is given by an integral over a set of
\(\Sigma\)'s that covers the moduli space \(\M\) of punctured super
Riemann surfaces. Let \(\sigma\) be a section of \(\P\) over \(\M\),
then
\begin{equation}
  \label{ampl}
  \la\!\la{\psi_1}\,\psi_2\cdots\psi_n\ra\!\ra = \int_{\sigma(\M)}
   \mu_\Psi.
\end{equation}
Where \(\mu_\Psi\) is the {\em superstring measure} given by the
following expression \cite{ag88-2}
\begin{equation}
  \label{smeasure}
  \mu_\Psi(V_F; \Sigma)=\bra\Sigma \dl^{d_e|d_o}
  (\,\la B, V_F\ra\,)\ket\Psi,
\end{equation}
where \(d_e|d_o\) is the dimension of the super moduli space and
\(F=1,\dots,d_e|d_o\). For genus \(g\) surfaces with \(p\)
Neveu-Schwarz and \(q\) Ramond punctures
\(d_e|d_o=3g-3+p+q|2g-2+p+{q\over2}\).  The pairing \(\la B, V_F\ra\)
is defined as in \eq{pairing} by
\begin{equation}
  \label{pair-b}
  \la B, V \ra=\sum_{\rm punctures}\oint \d z \d\th\,
  B^{(i)}(z,\th) V^{(i)}(z,\th);
\end{equation}
moreover, the two are related by \(\{Q,\la B, V_F\ra\} = \la\TT,
V_F\ra\).

Substituting the delta function by its integral representation we
immediately recognize a \(d_e|d_o\)-form which is a Baranov-Schwartz
transform of
\begin{equation}
  \label{psss}
  \omega_\Psi(\Sigma,\d\Sigma)=\bra\Sigma e^{i\la B,\d\Sigma\ra}\ket\Psi;
\end{equation}
in other words,
\begin{equation}
  \label{medif}
  \mu_\Psi=\lm^{d_e|d_o}\,\omega_\Psi.
\end{equation}

The main property of \(\omega_\Psi\) which follows easily from the
definitions is that
\begin{equation}
  \label{main}
  \d\,\omega_\Psi = \omega_{Q \Psi},
\end{equation}
This property implies immediately that the
string amplitude (\ref{ampl}) does not depend on \(\sigma\) or on the
choice of local coordinates when the scattering states are BRST
physical, and also that it does not change when the states are changed
by adding BRST trivial vectors:
\begin{equation}
  \label{sind}
  \int_{\sigma(\M)}\mu_\Psi = \int_{\sigma'(\M)}\mu_\Psi,
\end{equation}
and
\begin{equation}
  \label{qind}
    \int_{\sigma(\M)} \mu_\Psi = \int_{\sigma(\M)}\mu_{\Psi + Q\Lambda}.
\end{equation}
Both eqs.~(\ref{sind}) and (\ref{qind}) are direct consequences of the main
identity (\ref{main}) and the Stokes theorem. Another way to interpret
\eq{main} is to say that eqs.~(\ref{psss}) and (\ref{medif}) define a 
natural map from BRST cohomology to  de Rham cohomology of \(\P\).

\section{Acknowledgments}
I would like to thank L.~Dolan and B.~Zwiebach for valuable comments.
I am especially grateful to Th.~Voronov for the enlightening e-mail
communication we had during the last summer. 


\end{document}